\documentclass[prl,english,preprintnumbers,amsmath,amssymb,nofootinbib,twocolumn,superscriptaddress]{revtex4-1}
\usepackage[utf8]{inputenc}
\usepackage[T1]{fontenc}
\usepackage{amsmath}
\usepackage{amssymb}
\usepackage{graphicx}
\usepackage{physics}
\usepackage{tabularx}
\usepackage{bm}
\usepackage{pifont}

\usepackage{color}
\definecolor{orange}{rgb}{1,0.5,0}

\definecolor{bblue}{rgb}{0.2,0.8,1.0}

\newcommand{\beq}{\begin{equation}}
\newcommand{\eeq}{\end{equation}}
\newcommand{\bea}{\begin{eqnarray}}
\newcommand{\eea}{\end{eqnarray}}

\begin{document}

\title{The fourth- and fifth-order virial coefficients from weak-coupling to unitarity}

\author{Y. Hou}
\affiliation{Department of Physics and Astronomy, University of North Carolina, Chapel Hill, North Carolina 27599, USA}

\author{J. E. Drut}
\affiliation{Department of Physics and Astronomy, University of North Carolina, Chapel Hill, North Carolina 27599, USA}

\date{\today}

\begin{abstract}
In the current era of precision quantum many-body physics, one of the most scrutinized systems is the unitary limit of the 
nonrelativistic spin-$1/2$ Fermi gas, due to its simplicity and relevance for atomic, condensed matter, and nuclear physics. The 
thermodynamics of this strongly correlated system is determined by universal functions which, at high temperature, are
governed by universal virial coefficients $b_n$ that capture the effects of the 
$n$-body system on the many-body dynamics. Currently, $b_2$ and $b_3$ are well understood, but the situation is less clear for $b_4$,
and no predictions have been made for $b_5$.
To answer these open questions, we implement a nonperturbative analytic approach based on the Trotter-Suzuki factorization of the 
imaginary-time evolution operator, using progressively finer temporal lattice spacings. Implementing these factorizations and automated 
algebra codes, we obtain the interaction-induced change $\Delta b_n$ from weak coupling to unitarity. At unitarity, we find: 
$\Delta b_3 = -0.356(4)$, in agreement with previous results; $\Delta b_4 = 0.062(2)$, in agreement with all previous theoretical 
estimates but at odds with experimental determinations; and $\Delta b_5 = 0.078(6)$, which is a prediction. We show the impact of those answers on the density equation of state and Tan contact, and track their origin back to their polarized and unpolarized components.
\end{abstract}

\maketitle 

{\it Introduction.-} 
With the advances in precise ultracold atom experiments on one hand, and
new and powerful algorithms and machines on the other, quantum many-body physics has 
in many ways entered a precision era.
Experimentally, ultracold atoms are arguably the cleanest and most malleable 
systems~\cite{UltracoldAtoms1}, and also those where an ever-increasing number of observables can be measured 
with unprecedented precision~\cite{UltracoldAtoms4}.
On the computational side, progress has been steady in a wide range of areas: 
from advanced benchmarks of paradigmatic condensed matter systems like the Hubbard model~\cite{HubbardBenchmark} to 
percent-level calculations in lattice QCD~\cite{latticeQCDAmy}.

In this broad quantum many-body context, one of the most sought-after systems, due to its relevance to atomic, condensed matter, and nuclear physics, is the unitary limit of the three-dimensional spin-$1/2$ Fermi gas~\cite{ZwergerBook}. 
This system is remarkable for its deceptive simplicity:
it is just a two-species Fermi gas with an attractive zero-range interaction, tuned
to the threshold of two-body bound-state formation (i.e. infinite scattering length). 
While simple to define, the problem is challenging for many-body theory, as there are no small parameters
to perform a controlled expansion. In nature, the system is realized approximately in dilute neutron matter in the crust of neutron stars~\cite{PethickRavenhall} and practically exactly in
ultracold-atom experiments~\cite{UltracoldAtoms4}. In the latter, Feshbach resonances enable varying
the coupling strength by dialing an external magnetic field~\cite{Feshbach}, such that a large swath of the so-called
BCS-BEC crossover (which contains the unitary limit) can be realized and explored~\cite{UltracoldAtoms2, UltracoldAtoms3,StrinatiReview}.
The strongly coupled region around unitarity
is also interesting due to its strong pairing correlations, which modify both the superfluid
phase as well as the normal phase~\cite{Pseudogap, Pseudogap2}. Crucially, the unitary limit features
a nonrelativistic conformal invariance~\cite{UFGConformal} which is responsible for its hallmark property of 
{\it universality}~\cite{Universality}: it is characterized by dimensionless functions that are insensitive to the 
details of the underlying interactions.

Experiments realizing the unitary Fermi gas can achieve temperatures low-enough to probe the 
superfluid state, but also high-enough to access the normal state and the quantum-classical crossover. 
The latter is also of relevance to nuclear astrophysics~\cite{SchwenkHorowitz1,SchwenkHorowitz2,SchwenkHorowitz3}
and is characterized by the virial expansion~\cite{VirialReview}, whose behavior is determined by
{\it universal} virial coefficients $b_n$. At $n$-th order, these coefficients capture the thermodynamic contributions 
of the $n$-body system. While the calculation of $b_n$ of noninteracting gases is a textbook example, the 
interacting counterpart poses a challenging problem, especially so as $n$ is increased beyond $n = 2$.
The interaction-induced change in 
the second-order coefficient is set by the celebrated Beth-Uhlenbeck (BU) formula~\cite{BU,LeeSchaeferPRC1}
\bea
\label{Eq:BU}
\Delta b_2 =
		\frac{e^{\lambda^2}}{\sqrt{2}} \left [1 + \mbox{erf}(\lambda) \right], 
\eea
where $\lambda = \sqrt{\beta}/a_0$, $a_0$ is the s-wave scattering length, and $\beta$ is the inverse temperature. 
[The unitary limit corresponds to $\lambda = 0$; in this work 
we will focus on the $\lambda \leq 0$ sector.]
Meanwhile, the third-order coefficient $\Delta b_3$ is much more challenging to compute and 
has been approached numerically~\cite{LiuHuDrummond} as well as analytically~\cite{BedaqueRupak, DBK, Leyronas, CastinWerner,GaoEndoCastin}, at and away from unitarity. In turn, work on $\Delta b_4$ has largely focused 
on the unitary limit~\cite{Rakshit,EndoCastin,YanBlume,Castin} (see however Ref.~\cite{Ngampruetikorn}).
Notably, while there is good agreement on $\Delta b_3$ between theory and experiment~\cite{Nascimbene, Exp2},
the situation is less clear for $\Delta b_4$, as we explain below. There have been no estimates of $\Delta b_5$,
to the best of our knowledge.

In this work, we contribute to the exploration of the quantum-classical crossover by calculating
$\Delta b_4$ and $\Delta b_5$ (along with their counterparts for polarized systems) for spin-$1/2$ fermions 
with attractive interactions, covering from weak coupling to the unitary limit. To that end, we implement and 
progressively refine a factorization of the Boltzmann weight, extrapolating to the continuum limit 
of that factorization at the end. Our method is similar to the one originally 
advocated in Ref.~\cite{HouEtAl} but with dramatic improvements and optimizations that enabled
the present work. Below, we outline the formalism and basic aspects of the method, leaving the most 
technical details for the Supplemental Materials~\cite{SupMat}.

{\it Hamiltonian, virial expansion, and computational method.-} 
The Hamiltonian that describes dilute, two-species Fermi gases is 
$\hat H = \hat T + \hat V$, where
\bea
\label{Eq:T}
\hat T \!=\! \sum_{s=\uparrow,\downarrow}{\int{d^3 x\,\hat{\psi}^{\dagger}_{s}({\bf x})\left(-\frac{\hbar^2\nabla^2}{2m}\right)\hat{\psi}_{s}({\bf x})}},
\eea
and
\bea
\label{Eq:V}
\hat V \!=\! - g \! \int{d^3 x\,\hat{n}_{\uparrow}({\bf x})\hat{n}_{\downarrow}({\bf x})},
\eea
where $\hat{\psi}_{s}, \hat{\psi}^{\dagger}_{s}$ are the fermionic field operators for particles of spin $s= \uparrow,\downarrow$, and 
$\hat{n}_{s}({\bf x})$ are the coordinate-space densities. In the remainder of this work, we will take 
$\hbar = k_\text{B} = m = 1$. The contact interaction is singular in three-dimensional space,
such that regularization and renormalization are needed (see below).

The virial expansion is an expansion of the equilibrium many-body problem around the dilute limit
$z\to 0$, where $z=e^{\beta \mu}$ is the fugacity and $\mu$ the chemical potential coupled 
to the total particle number operator $\hat N$.
In powers of $z$, the grand-canonical partition function is
\beq
\label{Eq:Z}
\mathcal Z = \tr \left[e^{-\beta (\hat H - \mu \hat N)}\right] = \sum_{N=0}^{\infty} z^N Q_N,
\eeq 
where $Q_N = \tr_N\left[\exp(-\beta \hat H) \right]$ is the $N$-body partition function. 
Calling $\mathcal Z_0$ the noninteracting limit of $\mathcal Z$,
\beq
\ln \left({\mathcal Z}/ {\mathcal Z}_0 \right) = Q_1 \sum_{n=2}^{\infty} \Delta b_n z^n,
\eeq
where the interaction-induced change $\Delta b_n$ is related to the interaction change $\Delta Q_N$
by Taylor-expanding the logarithm of Eq.~(\ref{Eq:Z}) (and its noninteracting counterpart) around $z=0$
(see e.g. Ref.~\cite{HouEtAl} for explicit formulas).
To evaluate the $\Delta Q_{N}$ relevant for $\Delta b_n$, we
introduce a Trotter-Suzuki (TS) factorization of the imaginary-time evolution operator
\beq
\label{Eq:TSFactorization}
e^{-\beta \hat H} \simeq \left( e^{-\beta \hat T /(2 k)}e^{-\beta \hat V/ k} e^{-\beta \hat T /(2 k)} \right)^k,
\eeq
which defines the $k$-th order in the proposed approximation. 
When calculating $Q_N$, Eq.~(\ref{Eq:TSFactorization}) will appear inside a trace, such that the 
remainder scales as $\sim k^{-2}$.
Our code calculates symbolically the diagonal matrix elements of the right-hand side of Eq.~(\ref{Eq:TSFactorization})
in a complete set of Slater determinant multi-particle states (built out of plane-wave single-particle states),
inserting completeness relations between kinetic- and potential-energy factors.
The resulting momentum sums become Gaussian integrals in the continuum limit (see below), which 
are carried out analytically and automatically.
Previous work carried out calculations at $k=1$ for $\Delta b_n$ up to $n=4$
(the so-called semiclassical approximation of Refs.~\cite{ShillDrut, MorrellEtAl, HouEtAl}); 
and $k=2$ up to $n=7$~\cite{HouEtAl}. For the calculations
presented here, we dramatically improved our implementation, extending our analysis 
of $n=3, 4, 5$ for $k$ as large as possible with the computational resources available to us, respectively $k=21, 12, 9$.

{\it Renormalization.-}
To renormalize the contact interaction, we implemented two different procedures, 
both of them regularized by placing the system on a spatial lattice of spacing $\ell$, which is
implicitly taken to zero at the end of the calculation by replacing momentum sums with integrals
from $-\infty$ to $\infty$.
These renormalization procedures, which yield consistent results at large-enough order $k$
in Eq.~(\ref{Eq:TSFactorization}) (see also~\cite{SupMat}), are as follows. 

The first way is to tune the lattice theory so that the order-$k$ factorized calculation of $\Delta b_2$, 
matches the continuum value set by the BU formula~Eq.~(\ref{Eq:BU})
(e.g. $\Delta b_2 = 1/\sqrt{2}$ at unitarity). To that end, we tune the dimensionless coupling
${\tilde C}$ (see~\cite{SupMat})
to reproduce the desired dimensionless value of $\Delta b_2$. 
This is the same renormalization procedure used in Refs.~\cite{ShillDrut, MorrellEtAl, HouEtAl} and it amounts 
to following the ``line of constant $\Delta b_2$'' as $k$ is varied.

The second way to renormalize is more conventional: at a given factorization order $k$, we tune the coupling so as to reproduce 
the largest eigenvalue of the exact two-body transfer matrix, namely $\exp(-\beta E_0 / k)$, where $E_0$ is the exact two-body 
ground-state energy. The matrix elements of the order-$k$ factorized transfer matrix can be easily computed, in particular in the 
center-of-mass frame. Using those matrix elements, it is easy to see~\cite{SupMat} that the desired $\tilde{C}$ is given by
\begin{equation}
 \tilde{C} = \frac{1}{(2 \pi x)^{\frac{3}{2}}}\lim_{\Lambda \to \infty} \left[ \sum^\Lambda_{\bf a} \frac{1}{\exp(- \frac{4 \pi^2 x}{k} (\eta_0^2 - {\bf a}^2) ) - 1} \right]^{-1},
\end{equation}
where $x = {\beta}/{L^2}$, $L$ is the size of system, and $\bf a$ is a three-component integer vector.
Here, $\eta_0^2$ is set by the ground-state energy and given by L\"uscher's formula~\cite{Luescher}; 
at unitarity $\eta_0^2 \simeq -0.0959$. Taking the continuum limit amounts to opening the length scale 
window $\ell \ll \lambda_T \ll L$ (where $\lambda_T= \sqrt{2 \pi \beta}$ is the thermal wavelength),
which corresponds to calculating $\tilde{C}$ in the limit of 
$\Lambda\!\to\!\infty$ and $x\!\to\! 0$. 
This procedure follows the ``line of constant $E_0$'' as $k$ is varied. 
Since $\Delta b_2$ is sensitive to the whole energy spectrum, not just the ground-state energy $E_0$, the two procedures
yield answers which must be consistent at large enough $k$ if the continuum limit is approached.
We show our consistency checks in~\cite{SupMat}.

\begin{figure}[t]
  \begin{center}
  \includegraphics[scale=0.58]{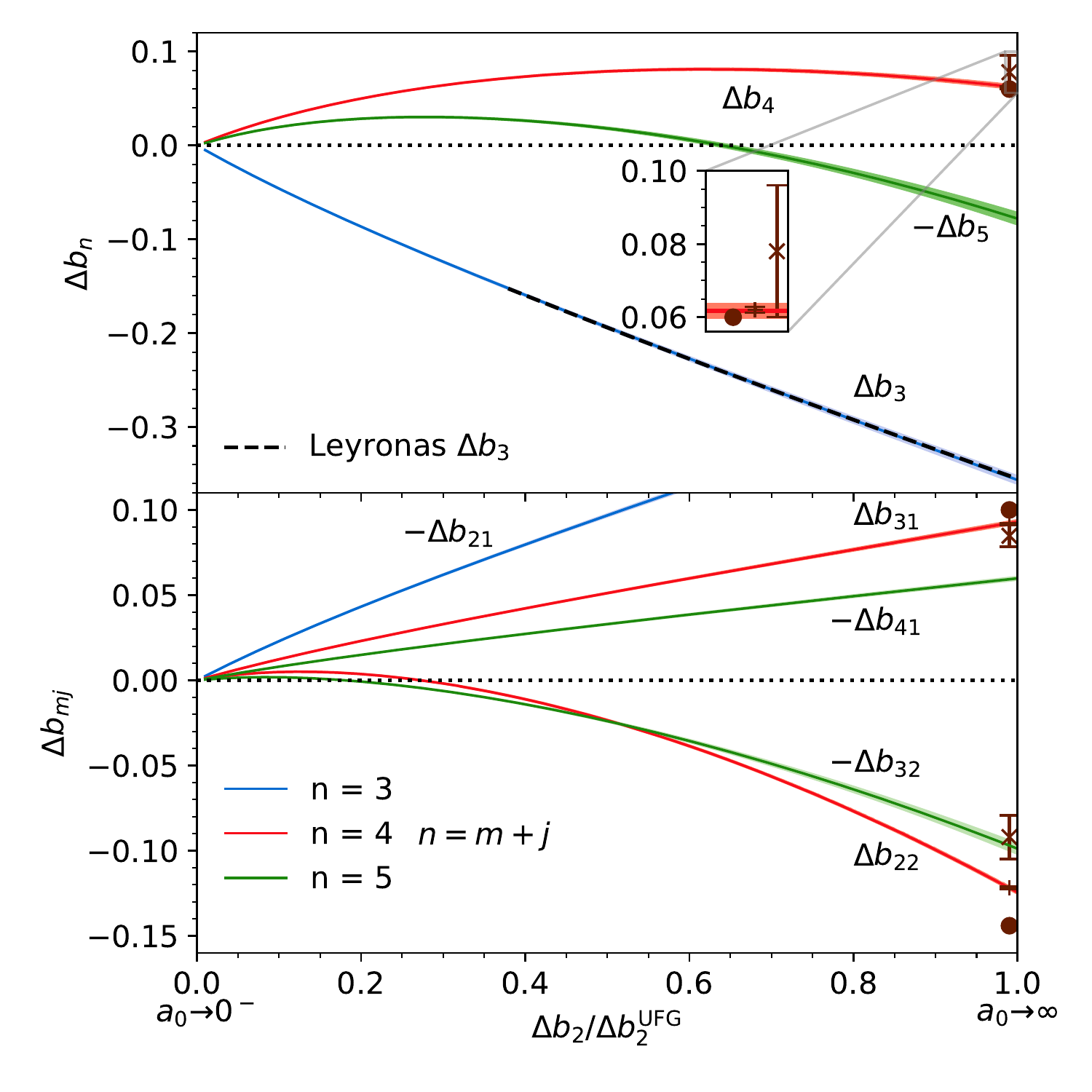}
  \end{center}
  \vspace{-0.5cm}
  \caption{{\bf Top}: Our results for
  \( \Delta b_3 \) (blue), \( \Delta b_4 \) (red) and \( -\Delta b_5 \) (green) shown with error bands 
  as functions of the coupling strength given by \( \Delta b_2 / \Delta b_2^{\mathrm{UFG}} \). 
  [We plot \( -\Delta b_5 \) to avoid display interference with \( \Delta b_4 \) around unitarity.]
  The dashed line shows $\Delta b_3$ from Ref.~\cite{Leyronas}.
  The dark red cross (with errorbar) shows the Monte Carlo results of Ref.~\cite{YanBlume}: \( \Delta b^\text{UFG}_{4} = 0.078(18)\); 
  the dark red plus sign (with small error bar) indicates the conjecture of Ref.~\cite{Castin}: \( \Delta b^\text{UFG}_{4} = 0.0620(8) \);
  finally, the dark red dot shows the approximate results of Ref.~\cite{Ngampruetikorn}: \( \Delta b^\text{UFG}_{4} = 0.06 \).
  {\bf Bottom}: Subspace contributions \( \Delta b_{mj} \) as functions of the coupling strength.
 Our results are shown as error bands, color-coded as in the top plot by $n = m+j$: 
 blue for $\Delta b_{21}$, red for $\Delta b_{31}$ and $\Delta b_{22}$, and green for 
 $\Delta b_{41}$ and $\Delta b_{32}$.
  The red cross (with errorbar) shows Ref.~: \( \Delta b^\text{UFG}_{31} = 0.0848(64)\) and \( \Delta b^\text{UFG}_{22} = -0.0920(128)\); the red dot shows Ref.~\cite{Ngampruetikorn}: \( \Delta b^\text{UFG}_{31} = 0.100 \) and \( \Delta b^\text{UFG}_{22} = -0.144 \); finally, the red plus sign (with small error bar) shows Ref.~\cite{Castin}: \( \Delta b^\text{UFG}_{31} = 0.09188(16) \) and \( \Delta b^\text{UFG}_{22} = -0.1216(8)\). Our results are closest to the latter; we obtain
  \( \Delta b^\text{UFG}_{31} = 0.0931(8) \) and \( \Delta b^\text{UFG}_{22} = -0.1244(7) \).
  } 
  \label{Fig:db4-vs-db2fraction-polarization}
\end{figure}

{\it Results.-} Using the methods described above, 
we obtained estimates for $\Delta b_n$, for $n=3,4,5$ by
extrapolating to the large-$k$ limit, with uncertainties in our answers resulting from that
extrapolation~\cite{SupMat}.
The results are shown in Fig.~\ref{Fig:db4-vs-db2fraction-polarization},
where we parametrize the coupling strength using the ratio 
$\Delta b_2/\Delta b_2^\text{UFG}$, where $\Delta b_2^\text{UFG} = 1/\sqrt{2}$ 
is the value of $\Delta b_2$ at unitarity. [The corresponding scattering length can be obtained via Eq.~(\ref{Eq:BU}).] 

As mentioned above, $\Delta b_3$ was estimated numerically as well as
(semi-)analytically by several authors~\cite{LiuHuDrummond, DBK, Leyronas, CastinWerner,GaoEndoCastin}, 
and is by now a well-understood number; at unitarity it is $\Delta b_3^\text{UFG} \simeq -0.3551$ 
(we quote only the first few digits of the exact-diagonalization result of 
Ref.~\cite{LiuHuDrummond}, which is enough for our needs here). We obtain 
$\Delta b^\text{UFG}_3 = -0.356(4)$ which, while not as precise as previous determinations, it agrees with them.
Our results are also in excellent agreement with Leyronas' analytic result~\cite{Leyronas}.

Also shown in Fig.~\ref{Fig:db4-vs-db2fraction-polarization} (top) are our results for $\Delta b_4$,
compared with prior theoretical estimates of its value at unitarity $\Delta b^\text{UFG}_4$.
Our result $\Delta b^\text{UFG}_4 = 0.062(2)$ compares well with every other theoretical estimate, namely
Yan and Blume~\cite{YanBlume}: \( \Delta b^\text{UFG}_{4} = 0.078(18)\); Endo and Castin~\cite{Castin}: \( \Delta b^\text{UFG}_{4} = 0.0620(8) \), and Ngampruetikorn et al.~\cite{Ngampruetikorn}: \( \Delta b^\text{UFG}_{4} = 0.06 \).
On one hand, the last two are a conjecture and an approximate result, respectively. On the other hand,
the first one is a Monte Carlo result which comes with a comparatively large uncertainty that encompasses all prior theoretical estimates.
Our calculation, like the Monte Carlo result, comes from a first-principles nonperturbative approach and agrees 
with all of the above results. However, our determination does not incur statistical errors and thus provides 
a substantial reduction in the overall uncertainty.

\begin{figure}[t]
  \begin{center}
  \includegraphics[scale=0.58]{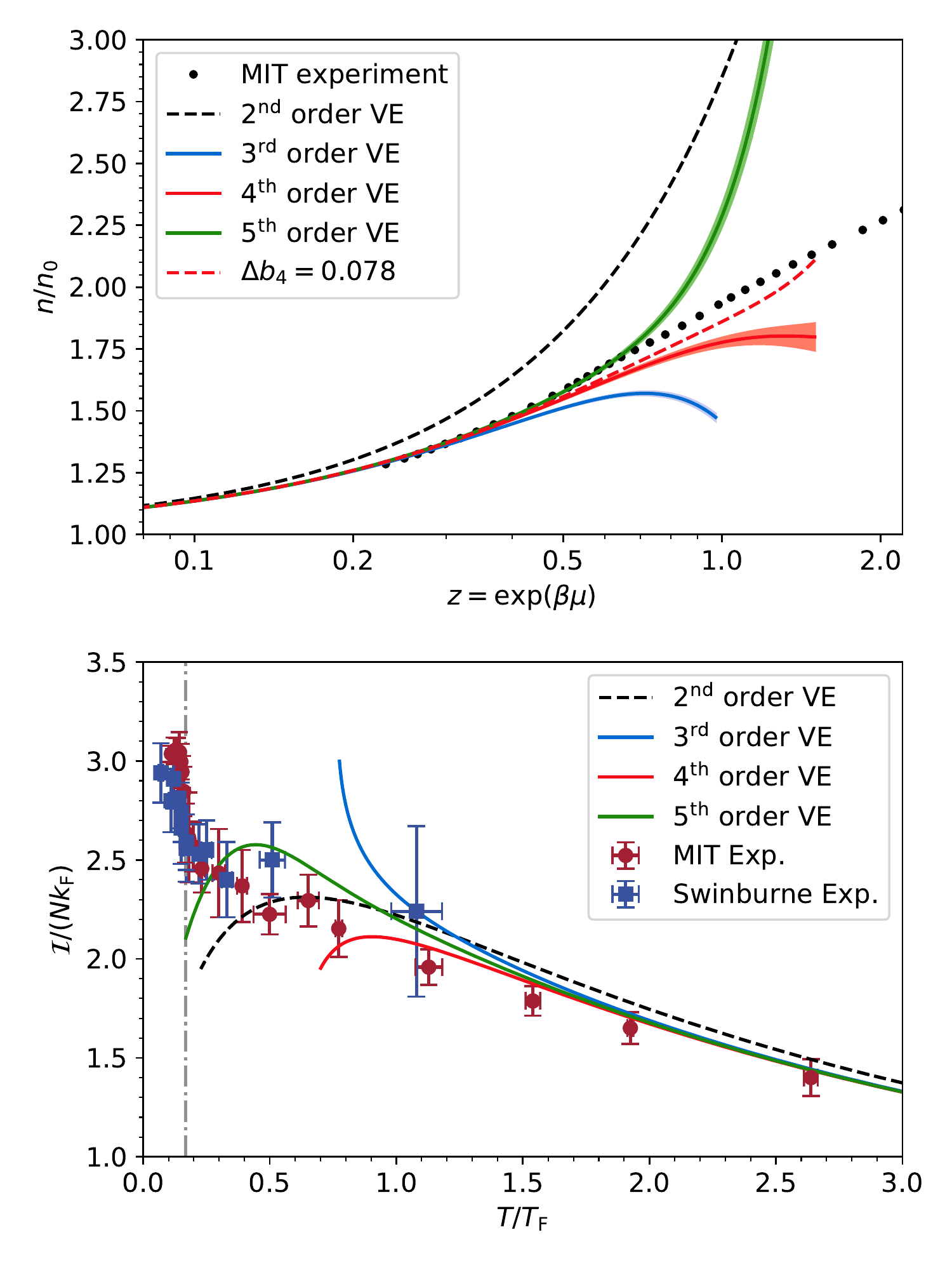}
  \end{center}
     \vspace{-0.2cm}
  \caption{{\bf Top}: Density equation of state at unitarity as a function of the fugacity $z$ showing our virial expansion results (error bands) 
  compared with the data of Ref.~\cite{Exp2}. The fourth-order virial expansion is also shown using the central value for $
  \Delta b_4$ of Ref.~\cite{YanBlume}.
  {\bf Bottom}: Tan contact at unitarity as a function of temperature $T$ in units of the Fermi temperature $T_F = (3\pi^2 n)^{2/3}/2$, 
  where $n$ is the density, compared with the experimental measurements of Refs.~\cite{SwinburneExp, MITExp}.
  The vertical dash-dotted line shows the critical temperature $T_c/T_F = 0.167(13)$ of Ref.~\cite{Exp2}.
  }
  \label{Fig:density-eos}
\end{figure}

There have also been attempts to determine $\Delta b^\text{UFG}_4$ 
from experimental data on the equation of state [e.g. from ENS \cite{Nascimbene}: 0.096(15), and MIT \cite{Exp2}: 0.096(10);
see also our Fig.~\ref{Fig:density-eos} (top)].
However, those analyses face a challenging numerical problem, namely fitting a fourth-order polynomial with no knowledge
of the size of higher order contributions or where the fourth order truly dominates; we return to this issue below.
While the Monte Carlo result of Ref.~\cite{YanBlume} overlaps with the above analyses, our 
result disagrees with them (as do Refs.~\cite{Castin,Ngampruetikorn}).

To understand the origin of our $\Delta b^\text{UFG}_4$,
we refer to Fig.~\ref{Fig:db4-vs-db2fraction-polarization} (bottom), which shows the two components that make up
the full result $\Delta b_4 = 2 \Delta b_{31} + \Delta b_{22}$: the polarized sector $\Delta b_{31}$ and the unpolarized sector
$\Delta b_{22}$ (defined in~\cite{SupMat}).
For essentially all couplings studied here, $\Delta b_{31}$ is increasingly positive and
$\Delta b_{22}$ increasingly negative as the unitary limit is approached. That competition results in the 
non-monotonic behavior of $\Delta b_4$ in Fig.~\ref{Fig:db4-vs-db2fraction-polarization} (top),
and in particular in its low value at unitarity.

For the fifth-order virial coefficient at unitarity we obtain $\Delta b^\text{UFG}_5 = 0.078(6)$, which
is the first estimate of this universal quantity, to the best of our knowledge. Figure~\ref{Fig:db4-vs-db2fraction-polarization} (top)
shows $\Delta b_5$ as a function of the coupling.
As with $\Delta b_4$, the non-monotonicity of $\Delta b_5$ can be traced back to the competition between
two sectors with (largely) monotonic but opposite behavior. As shown in Fig.~\ref{Fig:db4-vs-db2fraction-polarization} (bottom), 
$\Delta b_{41}$ and $\Delta b_{32}$ become progressively more negative and more positive, respectively,
as the coupling is increased (with the exception of a small region at very weak couplings where $\Delta b_{32}$
is negative). Thus, $\Delta b_5 = 2 \Delta b_{41} + 2\Delta b_{32}$ is non-monotonic;
furthermore, it changes sign from negative to positive around $\Delta b_2/\Delta b_2^\text{UFG} \simeq 0.63$ and
proceeds to grow in magnitude enough to overtake $\Delta b_4$. This is notable because the ``normal'' ordering 
$|\Delta b_3| > |\Delta b_4| > |\Delta b_5|$ is preserved from weak couplings up to $\Delta b_2/\Delta b_2^\text{UFG} \simeq 0.96$,
but $|\Delta b_5| > |\Delta b_4|$ after that, in particular at unitarity.
Crucially, such a large $\Delta b_5$ could easily interfere with the experimental determination of $\Delta b_4$,
which would explain the discrepancies between our results and the
experimental equation-of-state analyses of $\Delta b_4$. 

The subspace contributions $\Delta b_{mj}$ mentioned above
allow us to study the first steps of the ``polaron sequence'' 
$\Delta b_{m1}$. Beyond the qualitative resemblance of $|\Delta b_{m1}|$ for all $m$, 
we find that $|\Delta b_{m1}|$ decreases as $m$ is increased
for all the couplings we studied (see Table~\ref{Table:Summary} in particular), 
which we interpret as due to the largely noninteracting majority particles
(as the interaction is of zero range). Furthermore, we observe that the
sequence alternates in sign, which we conjecture will persist for arbitrary $m$.
\begin{table}[t]
  \normalsize
    \centering
    \caption{
    \label{Table:Summary} Estimates for $\Delta b_3$ to $\Delta b_5$ in the unitary limit, including
    the subspace coefficients for the polarized case $\Delta b_{mj}$.
    }
    \begin{tabular*}{\columnwidth}{@{\extracolsep{\fill}}c|cccc@{}}
      \hline\hline
                          &    $n = 3$ &  $n = 4$  & $n = 5$     \\
      \hline\hline
      \( \Delta b^\text{UFG}_n \)    &     -0.356(4)   &     0.062(2) &     0.078(6)       \\
      \hline
      \( \Delta b^\text{UFG}_{(n-1)1} \) &     -0.178(2)   &     0.0931(8) &       -0.0598(7)  \\
      \( \Delta b^\text{UFG}_{(n-2)2} \) &     --          &    -0.1244(7) &        0.0988(29) \\
      \hline\hline
    \end{tabular*}
\end{table}

In a harmonic trapping potential of frequency $\omega$, the $b_n$ (being
dimensionless) acquire a dependence on $\beta \omega$. In the high-temperature limit $\beta \omega \to 0$, the relationship
$\Delta b^\text{T}_n(\beta \omega) \to \Delta b^\text{T}_n = \Delta b_n/n^{3/2}$ holds~\cite{LiuHuDrummond}, where 
$\Delta b^\text{T}_n(\beta \omega)$ is the trapped coefficient, and $\Delta b^\text{T}_n$ its
high-temperature limit. At unitarity, $\Delta b^\text{T}_2 = 1/4$, and we find
$\Delta b^\text{T}_3 = -0.0685(8)$, $\Delta b^\text{T}_4 =  0.00775(25)$, and $\Delta b^\text{T}_5 = 0.0070(5)$.
Notably, the factor $n^{-3/2}$ restores the ``normal order'' $ |\Delta b^\text{T}_3| > |\Delta b^\text{T}_4| > |\Delta b^\text{T}_5|$,
in contrast to the homogeneous case, supporting the notion that trapping potentials enhance the convergence
of the virial expansion~\cite{LiuHuDrummond}.

Finally, in Fig.~\ref{Fig:density-eos} (top) we use our results to obtain the density equation of state 
and compare with the experiment of Ref.~\cite{Exp2}.
While our results at fourth order are somewhat farther away from the data than those of 
Ref.~\cite{YanBlume}, the fifth-order contribution considerably improves the agreement for 
$z=0.5-0.73$. In Fig.~\ref{Fig:density-eos} (bottom) we compare our results for the Tan 
contact~\cite{Tan1,Tan2, Tan3} with the measurements of Refs.~\cite{SwinburneExp, MITExp}. For clarity, 
we only compare with experiments; we include theoretical 
approaches and the polarized case in~\cite{SupMat}. Our fifth-order results appear to follow the trend of the experimental data
for temperatures as low as $T/T_F \simeq 0.45$.

{\it Summary and Conclusions.-} In this work we have performed a fully nonperturbative
calculation of the fourth- and fifth-order virial coefficients of attractively interacting spin-$1/2$ fermions,
from weak coupling to the unitary limit.
To that end, we implemented a TS factorization of the imaginary-time 
evolution operator, using progressively finer temporal lattice spacings and extrapolating to the continuous-time limit. 
The traces of these factorizations were
calculated analytically using automated algebra to access the canonical partition functions that yield 
the interaction-induced changes $\Delta b_n$.
We found that the universal values at unitarity are $\Delta b^\text{UFG}_3 = -0.356(4)$, in agreement with previous 
calculations; $\Delta b^\text{UFG}_4 = 0.062(2)$, in agreement with previous theoretical estimates but at odds with 
experimental equation-of-state analyses; and finally $\Delta b^\text{UFG}_5 = 0.078(6)$, which is a prediction.
To elucidate the origin of these answers, we showed the subspace contributions 
$\Delta b_{31}$, $\Delta b_{22}$, $\Delta b_{41}$, and $\Delta b_{32}$. 
We found that these tend to grow in magnitude with the coupling strength 
but come with opposite sign and thus compete within $\Delta b_n$. The $\Delta b_{ij}$ are 
important as they govern the virial expansion of polarized gases (see e.g.~\cite{PolarizedUFG}). To
show the impact of our $\Delta b^\text{UFG}_4$ and $\Delta b^\text{UFG}_5$, we compared with the experimental
determination of the density equation of state and the Tan contact. Our answers yield an improvement over lower orders,
which is remarkable considering the size of the contributions and that the system is strongly correlated.

{\it Acknowledgments.-} 
We would like to thank J. Braun, A. Bulgac, H. Hu, L. Rammelm\"uller, A. Richie-Halford, C. Vale, 
and M. W. Zwierlein for comments on the manuscript.
We are also grateful to the authors of Refs.~\cite{Leyronas, SwinburneExp, MITExp} for kindly providing us with their data.
This material is based upon work supported by the National Science Foundation under Grant No.
PHY{1452635}.




\begin{thebibliography}{99}


\bibitem{UltracoldAtoms1}
M. Lewenstein, A. Sanpera, V. Ahufinger,
{\it Ultracold Atoms in Optical Lattices: Simulating Quantum Many-body Systems},
(Oxford University Press, New York, 2012)

\bibitem{UltracoldAtoms4}
\textit{Ultracold Fermi Gases}, 
Proceedings of the International School of Physics ``Enrico Fermi", Course CLXIV, 
Varenna, June 20 -- 30, 2006, 
M.~Inguscio, W.~Ketterle, C.~Salomon (Eds.) (IOS Press, Amsterdam, 2008).


\bibitem{HubbardBenchmark}
J. P. F. LeBlanc, A. E. Antipov, F. Becca, I. W. Bulik, G. Kin-Lic Chan, C.-M. Chung, Y. Deng, M. Ferrero, T. M. Henderson, C. A. Jim{\'e}nez-Hoyos, E. Kozik, X.-W. Liu, A. J. Millis, N. V. Prokof'ev, M. Qin, G. E. Scuseria, H. Shi, B. V. Svistunov, L. F. Tocchio, I. S. Tupitsyn, S. R. White, S. Zhang, B.-X. Zheng, Z. Zhu, and E. Gull,
{\it Solutions of the Two-Dimensional Hubbard Model: Benchmarks and Results from a Wide Range of Numerical Algorithms},
Phys. Rev. X {\bf 5}, 041041 (2015).


\bibitem{latticeQCDAmy}
C.C. Chang, A.N. Nicholson, E. Rinaldi, E. Berkowitz, N. Garron,
{\it A per-cent-level determination of the nucleon axial coupling from quantum chromodynamics},
Nature {\bf 558}, 91 (2018).


\bibitem{ZwergerBook}
{\it The BCS-BEC Crossover and the Unitary Fermi Gas}, 
edited by W. Zwerger (Springer-Verlag, Berlin, 2012).


\bibitem{PethickRavenhall}
C. J. Pethick and D. G. Ravenhall, 
{\it Matter at large neutron excess and the physics of neutron-star crusts},
Ann. Rev. Nucl. Part. Sci. {\bf 45}, 429 (1995).


\bibitem{Feshbach}
C. Chin, R. Grimm, P. Julienne, and E. Tiesinga,
{\it Feshbach resonances in ultracold gases},
Rev. Mod. Phys. {\bf 82}, 1225 (2010).


\bibitem{UltracoldAtoms2}
I. Bloch, J. Dalibard, W. Zwerger
{\it Many-Body Physics with Ultracold Gases},
Rev. Mod. Phys. {\bf 80}, 885 (2008).


\bibitem{UltracoldAtoms3}
S. Giorgini, L.P. Pitaevskii, S. Stringari,
{\it Theory of ultracold Fermi gases},
Rev. Mod. Phys. {\bf 80}, 1215 (2008).


\bibitem{StrinatiReview}
G. C. Strinati, P. Pieri, G. R{\"o}pke, P. Schuck, and M. Urban, 
{\it The BCS-BEC crossover: From ultra-cold fermi gases to nuclear systems}, 
Phys. Rep. {\bf 738}, 1 (2018).


\bibitem{Pseudogap}
E. J. Mueller, 
{\it Review of pseudogaps in strongly interacting fermi gases}, 
Rep. Prog. Phys. {\bf 80}, 104401 (2017).

\bibitem{Pseudogap2}
A. Richie-Halford, J. E. Drut, A. Bulgac,
{\it Emergence of a pseudogap in the BCS-BEC crossover},
arXiv:2004.05014.


\bibitem{UFGConformal}
Y. Nishida and D. T. Son, 
{\it Nonrelativistic conformal field theories},
Phys. Rev. D {\bf 76}, 086004 (2007).

\bibitem{Universality}
T. L. Ho,
{\it Universal Thermodynamics of Degenerate Quantum Gases in the Unitarity Limit},
Phys. Rev. Lett. {\bf 92}, 090402 (2004).

\bibitem{SchwenkHorowitz1}
C. J. Horowitz and A. Schwenk,
{\it The Virial equation of state of low-density neutron matter},
Phys. Lett. B {\bf 638}, 153 (2006).


\bibitem{SchwenkHorowitz2}
C. J. Horowitz and A. Schwenk,
{\it The Neutrino response of low-density neutron matter from the virial expansion},
Phys. Lett. B {\bf 642}, 326 (2006).


\bibitem{SchwenkHorowitz3}
C. J. Horowitz and A. Schwenk,
{\it Cluster formation and the virial equation of state of low-density nuclear matter},
Nucl. Phys. A {\bf 776}, 55 (2006).

\bibitem{VirialReview}
X.-J. Liu,
{\it Virial expansion for a strongly correlated Fermi system and its application to ultracold atomic Fermi gases},
Phys. Rep. {\bf 524}, 37 (2013).


\bibitem{BU}
E. Beth and G. E. Uhlenbeck, 
{\it The quantum theory of the non-ideal gas. II. Behaviour at low temperatures},
Physica (Utrecht) {\bf 4}, 915 (1937).


\bibitem{LeeSchaeferPRC1}
D. Lee and T. Sch\"afer,
{\it Cold dilute neutron matter on the lattice. I. Lattice virial coefficients and large scattering lengths},
Phys. Rev. C {\bf 73}, 015201 (2006).


\bibitem{LiuHuDrummond}
X.-J. Liu, H. Hu, and P. D. Drummond,
{\it Virial expansion for a strongly correlated Fermi gas},
Phys. Rev. Lett. {\bf 102}, 160401 (2009).


\bibitem{BedaqueRupak}
P. F. Bedaque and G. Rupak
{\it Dilute resonating gases and the third virial coefficient},
Phys. Rev. B {\bf 67}, 174513 (2003).


\bibitem{DBK}
D. B. Kaplan, S. Sun,
{\it A new field theoretic method for the virial expansion},
Phys. Rev. Lett. {\bf 107}, 030601 (2011).


\bibitem{Leyronas}
X. Leyronas,
{\it Virial expansion with Feynman diagrams},
Phys. Rev. A {\bf 84}, 053633 (2011).


\bibitem{CastinWerner}
Y.Castin and F. Werner,
{\it Le troisi\'eme coefficient du viriel du gaz de Bose unitaire}, 
Can. J. Phys. {\bf 91}, 382 (2013).


\bibitem{GaoEndoCastin}
C. Gao, S. Endo, and Y. Castin,
{\it The third virial coefficient of a two-component unitary Fermi gas across an Efimov-effect threshold}, 
Europhys. Lett. {\bf 109}, 16003 (2015).


\bibitem{Rakshit}
D. Rakshit, K. M. Daily, and D. Blume, 
{\it Natural and unnatural parity states of small trapped equal-mass two-component Fermi gases at unitarity and fourth-order virial coefficient},
Phys. Rev. A {\bf 85}, 033634 (2012).


\bibitem{EndoCastin}
S. Endo and Y. Castin,
{\it Absence of a four-body Efimov effect in the 2 + 2 fermionic problem},
Phys. Rev. A {\bf 92}, 053624 (2015).


\bibitem{YanBlume}
Y. Yan, D. Blume,
{\it Path integral Monte Carlo determination of the fourth-order virial coefficient for unitary two-component Fermi gas with zero-range interactions},
Phys. Rev. Lett. {\bf 116}, 230401 (2016).


\bibitem{Castin}
S. Endo and Y. Castin,
{\it The interaction-sensitive states of a trapped two-component ideal Fermi gas and application to the virial expansion of the unitary Fermi gas},
J. Phys. A {\bf 49}, 265301 (2016).

\bibitem{Ngampruetikorn}
V. Ngampruetikorn, M. M. Parish, and J. Levinsen,
{\it High-temperature limit of the resonant Fermi gas},
Phys. Rev. A {\bf 91}, 013606 (2015).

\bibitem{Nascimbene}
S. Nascimbene, N. Navon, K. Jiang, F. Chevy, and C. Salomon, 
{\it Exploring the thermodynamics of a universal Fermi gas}, 
Nature {\bf 463}, 1057 (2010).


\bibitem{Exp2} 
M. J. H. Ku, A. T. Sommer, L. W. Cheuk, and M. W. Zwierlein,
{\it Revealing the Superfluid Lambda Transition in the Universal Thermodynamics of a Unitary Fermi Gas},
Science {\bf 335}, 563 (2012).




\bibitem{HouEtAl}
Y. Hou, A. Czejdo, J. DeChant, C. R. Shill, J. E. Drut,
{\it Leading-order semiclassical approximation to the first seven virial coefficients of spin-1/2 fermions across spatial dimensions},
Phys. Rev. A {\bf 100}, 063627 (2019).


\bibitem{SupMat}
Supplemental Materials: Technical aspects of the method, extrapolations to the continuous time limit, and
results for physical quantities. See ancillary file in arXiv.


\bibitem{ShillDrut}
C. R. Shill, J. E. Drut,
{\it Virial coefficients of 1D and 2D Fermi gases by stochastic methods and a semiclassical lattice approximation},
Phys. Rev. A {\bf 98}, 053615 (2018).



\bibitem{MorrellEtAl}
K. J. Morrell, C. E. Berger, and J. E. Drut,
{\it Third- and fourth-order virial coefficients of harmonically trapped fermions in a semiclassical approximation},
Phys. Rev. A {\bf 100}, 063626 (2019).



\bibitem{Luescher}
M. L\"uscher, 
{\it Volume dependence of the energy spectrum in massive quantum field theories},
Commun.\ Math.\ Phys. {\bf 105}, 153 (1986). 

\bibitem{Tan1}
S. Tan, 
{\it Energetics of a strongly correlated Fermi gas},
Ann. Phys. {\bf 323}, 2952 (2008); 

\bibitem{Tan2}
S. Tan, 
{\it Large momentum part of fermions with large scattering length},
Ann. Phys. {\bf 323}, 2971 (2008); 

\bibitem{Tan3}
S. Tan, 
{\it Generalized Virial Theorem and Pressure Relation for a strongly correlated Fermi gas},
Ann. Phys. {\bf 323}, 2987 (2008).

\bibitem{SwinburneExp}
C. Carcy, S. Hoinka, M. G. Lingham, P. Dyke, C. C. N. Kuhn, H. Hu, and C. J. Vale,
{\it Contact and Sum Rules in a Near-Uniform Fermi Gas at Unitarity},
Phys. Rev. Lett. {\bf 122}, 203401 (2019).


\bibitem{MITExp}
B. Mukherjee, P. B. Patel, Z. Yan, R. J. Fletcher, J. Struck, and M. W. Zwierlein,
{\it Spectral Response and Contact of the Unitary Fermi Gas},
Phys. Rev. Lett. {\bf 122}, 203402 (2019).


\bibitem{PolarizedUFG}
L. Rammelm\"uller, A. C. Loheac, J. E. Drut, and J. Braun,
{\it Finite-Temperature Equation of State of Polarized Fermions at Unitarity},
Phys. Rev. Lett. {\bf 121}, 173001 (2018).


\end{thebibliography}
\end{document}